\documentclass{aa}  

\usepackage{graphicx}
\usepackage{txfonts}
\usepackage{soul}
\usepackage{float}
\usepackage[colorlinks,citecolor=blue,urlcolor=blue]{hyperref}
\usepackage{subcaption}

\begin{document}

   \title{The dynamics of small-scale magnetic fields modulated by the solar cycle}

   \author{M. Stangalini \inst{1}
          \and
          G. Verth \inst{2}
          \and
          V. Fedun \inst{3}
          \and
          D. Perrone \inst{1}
          \and
          M. Berretti \inst{4,5}
          \and
          P. Bushby \inst{6}
          \and
          S. Jafarzadeh \inst{7,8}
          \and
          D. B. Jess \inst{7,9}
          \and
          F. Giannattasio \inst{10}
          \and
          P. H. Keys \inst{7}
          \and
          R. Bruno \inst{11}
          \and
          F. Berrilli \inst{5}
          }

   \institute{ASI Italian Space Agency, Via del Politecnico snc, 00133 Rome, Italy
              \email{marco.stangalini@asi.it}
          \and
          Plasma Dynamics Group, School of Mathematical and Physical Sciences, The University of Sheffield, Hicks Building, Hounsfield Road, Sheffield, S3 7RH, UK
          \and
          Plasma Dynamics Group, School of Electrical and Electronic Engineering, The University of Sheffield, Mappin Street, Sheffield, S1 3JD, UK
          \and 
          University of Trento, Via Calepina 14, 38122 Trento, Italy
          \and
          University of Rome Tor Vergata, Department of Physics, Via della Ricerca Scientifica 3, 00133 Rome, Italy
          \and
          School of Mathematics, Statistics and Physics, Newcastle University, Newcastle upon Tyne, NE1 7RU, UK
          \and
          Astrophysics Research Centre, School of Mathematics and Physics, Queen’s University Belfast, Belfast, BT7 1NN, Northern Ireland, UK
          \and
          Niels Bohr International Academy, Niels Bohr Institute, Blegdamsvej 17, DK-2100 Copenhagen, Denmark
          \and
          Department of Physics and Astronomy, California State University Northridge, Northridge, CA 91330, USA
          \and
          INGV, Istituto Nazionale di Geofisica e Vulcanologia, Rome, Italy
          \and
          INAF-IAPS, Istituto Nazionale di Astrofisica, Rome, Italy}

   \date{}

  \abstract
{
    In addition to sunspots, which represent the most easily visualized manifestation of solar magnetism, cutting-edge observations of the solar atmosphere have uncovered a plethora of magnetic flux tubes, down to the resolving power of modern high-resolution telescopes (a few tens of km), revealing how the Sun is a fully magnetized star. These magnetic elements are advected and buffeted by ambient plasma flows and turbulent convection, resulting in perturbations of the flux tubes that make them natural conduits for channeling wave energy into the upper layers of the Sun’s atmosphere and significantly contributing to the acceleration of the solar wind. Today, data acquired by the Helioseismic and Magnetic Imager (HMI) onboard NASA's Solar Dynamics Observatory (SDO), have made it possible to study the dynamics of small-scale magnetic fields over long timescales. Here, for the first time, we present the discovery of a modulation in the dynamical behavior of small-scale magnetic concentrations in the photosphere over temporal scales consistent with the solar activity cycle (i.e., $11$ years), which has only been made possible by the long observing lifetime of the SDO/HMI spacecraft. Furthermore, a temporal varying polarization of their perturbations is also found on similar timescales. This demonstrates how the small-scale dynamics of magnetic fields are also affected by the global dynamo. These discoveries were realized through automated tracking of magnetic fields in the solar photosphere across $11$ continuous years, resulting in the most extended statistical analyses of its kind so far, with more than $31$~million magnetic concentrations examined.
   }

   \keywords{
               }

   \maketitle
%

\section{Introduction}

    The solar photosphere presents magnetic features over a wide range of scales, from scales typical of sunspots of tens of megameters, down to spatial scales close to the resolving power of modern high-resolution telescopes \citep[$50-100$~km,][]{2009SSRv..144..275D, 2010ApJ...723L.164L, 2012A&A...540A..66L}. Subject to the turbulent forcing of the ambient photospheric plasma, small-scale magnetic fields are advected and diffused over the solar surface \citep[][]{2011ApJ...743..133A, 2012ApJ...759L..17L, 2013ApJ...770L..36G, 2017ApJS..229....8J}. This forcing can also excite magnetohydrodynamic (MHD) waves \citep[][]{1983SoPh...88..179E, 2014A&A...569A.102S, 2016GMS...216..449J, 2023LRSP...20....1J}, which can propagate upward and take part in the energization of the outer layers of the solar atmosphere and thus in the acceleration of the solar wind \citep[][]{2003ApJ...585.1138H, 2008ApJ...680.1542H, 2010ApJ...710.1857M, 2017PJAB...93...87S, 2020SSRv..216..140V}.\\
    Based on high-resolution observations of the solar atmosphere, many authors have reported horizontal perturbations of small-scale magnetic elements with velocity amplitudes on the order of $1-2$~km/s \citep[][]{2011ApJ...740L..40K, 2013A&A...549A.116J, 2013A&A...554A.115S}, consistent with the horizontal velocity flows of the ambient photospheric plasma \citep[][]{2010ApJ...716L..19M}. However, these analyses were based on data sequences limited to $1-2$ hours maximum, thus not capturing a possible evolution of their dynamics over longer timescales.
    \begin{figure*}[t!]
       \centering
       \includegraphics[width=17cm,trim={13cm 0 0 0},clip]{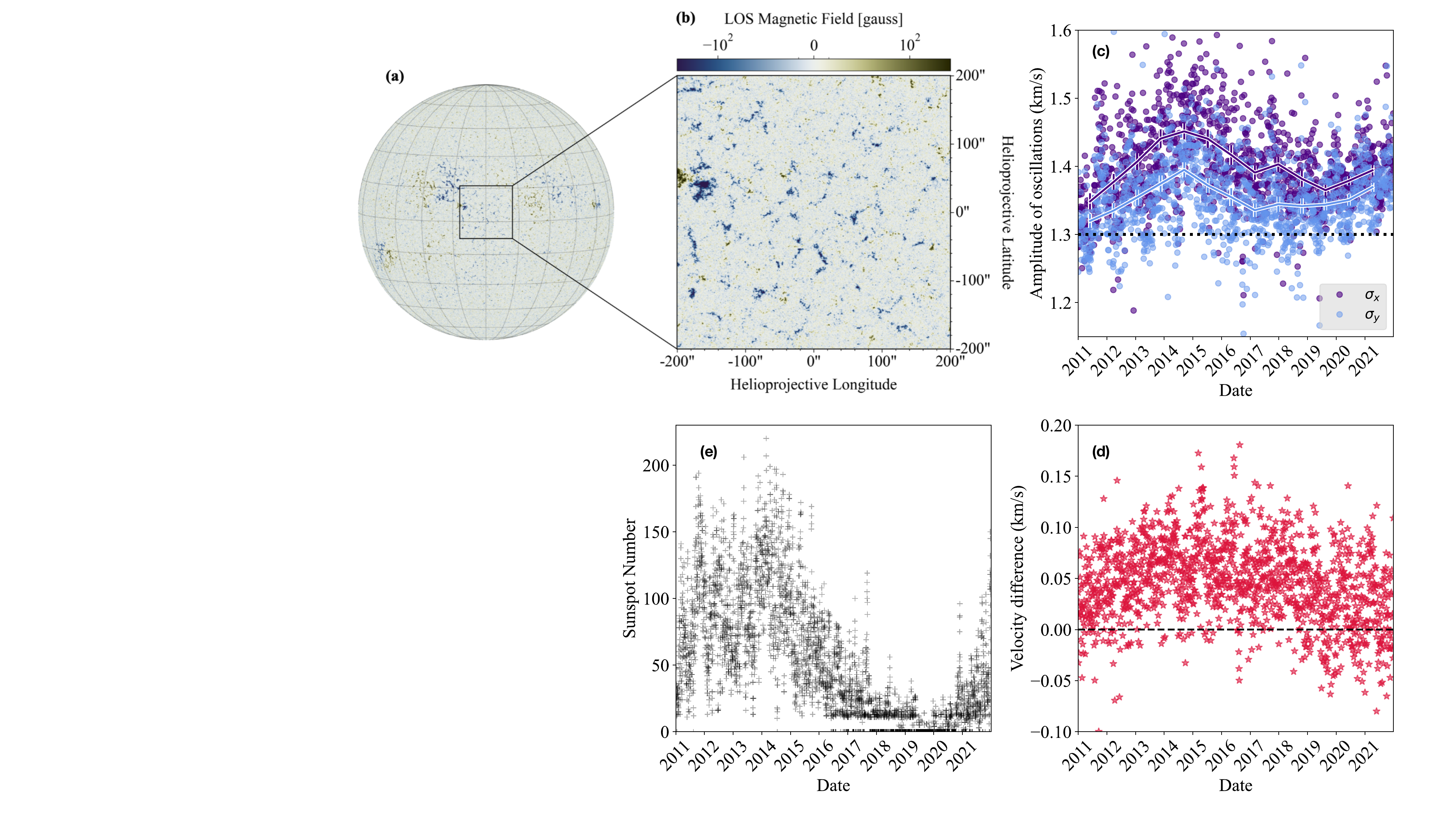}
       \caption{(a) Example of full-disk magnetogram acquired by SDO/HMI. (b) FoV considered in the analysis. Magnetic elements are automatically detected and tracked for 2 hours every three days in the period $2011-2022$. (c) Mean amplitude of the horizontal oscillations of the magnetic elements in the East-West (x) and North-South (y) directions. The error bars represent the standard error of the mean. (d) Difference of the amplitude in the two directions. (e) Sunspot number in the same period analyzed, indicating the progression of the solar cycle. The dynamics of small-scale and short-lived magnetic elements in the solar atmosphere is affected by a long term modulation on much longer timescales (years), compared to their lifetimes (several minutes). These dynamical properties cannot be explained only by a local driver of the perturbations, and indicates the presence of a global scale effect.}
       \label{fig:1}
   \end{figure*}

    Launched in $2012$ and with more than $10$ years of continuous and stable observations of the solar photospheric magnetic fields, the HMI magnetograph \citep[][]{2012SoPh..275..207S} onboard NASA’s SDO mission \citep[][]{2012SoPh..275....3P} enables the investigation of the dynamics of small-scale fields in the solar atmosphere over temporal scales typical of the solar activity cycle. This allows for the investigation of the dynamics of photospheric magnetic flux tubes and their possible link to global-scale phenomena such as the dynamo.
    We utilize here the unprecedented long-term observational capabilities of SDO/HMI to conduct a comprehensive study of small-scale magnetic concentration dynamics.  Over 31 million such features are analyzed across a complete solar cycle, enabling investigation of their relationship to global solar activity.

\section{Dataset}
    In this work, we used sequences of photospheric magnetograms obtained in the Fe~{\sc i}~617.3 nm spectral line and acquired by the Heliosesmic Magnetic Imager \citep[HMI][]{2012SoPh..275..207S} onboard the NASA Solar Dynamics Observatory \citep[SDO][]{2012SoPh..275....3P} with a cadence of 45 s.\\
    Specifically, magnetic flux tubes in a region $400\times400$~arcsec$^2$ ($290\times290$~Mm$^2$), located at the center of the solar disk, are tracked in HMI magnetograms (Fig.~\ref{fig:1} panels a and b) and their horizontal motions are analyzed by employing an automatic tracking algorithm. For more details about the data used and the tracking we refer the reader to \citet{2024A&A...687L..21B}. This analysis spans $11$ years, resulting in an accumulated set of more than $31$ million small-scale magnetic concentrations. More in particular, we consider data sequences of HMI magnetograms of 40 minutes every 3 days in the period 1 January 2011 – 31 December 2021, thus covering an entire solar activity cycle. The choice of the length of the data segments was made to allow the detection of the typical photospheric frequencies (i.e. $3$ mHz), while maintaining the data volume at a reasonable level. In this regard, it is also worth considering that the average lifetime of the tracked magnetic elements is $\sim 20$ min. Additionally, since solar rotation introduces a characteristic timescale of $\sim 27$ days, a three-day sampling ensures that different solar longitudes are systematically covered, mitigating potential aliasing effects.

    \begin{figure}[h!]
       \centering
       \includegraphics[width=7cm]{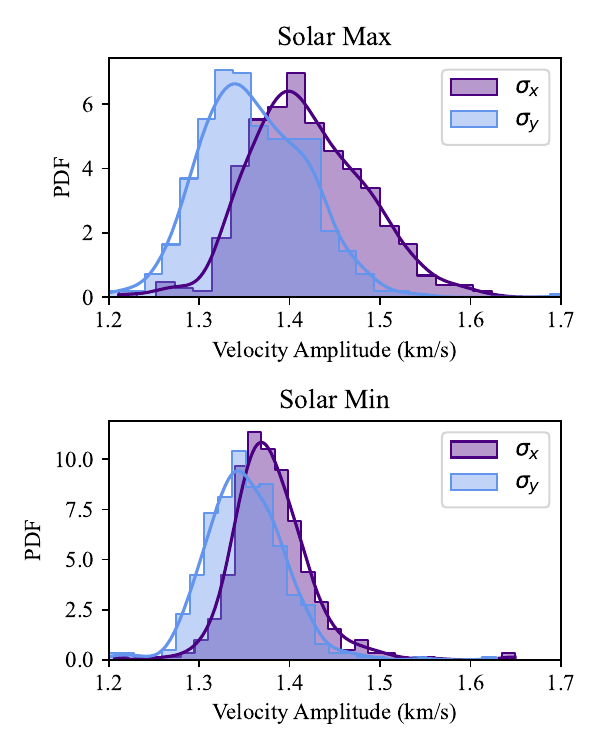}
       \caption{Probability Density Function (PDF) of the horizontal oscillations of the magnetic structures in both x and y direction at the solar maximum and at the solar minimum.}
       \label{fig:2}
   \end{figure}

     \begin{figure}
         \centering
         \begin{subfigure}[b]{0.4\textwidth}
           \centering
           \includegraphics[width=\linewidth]{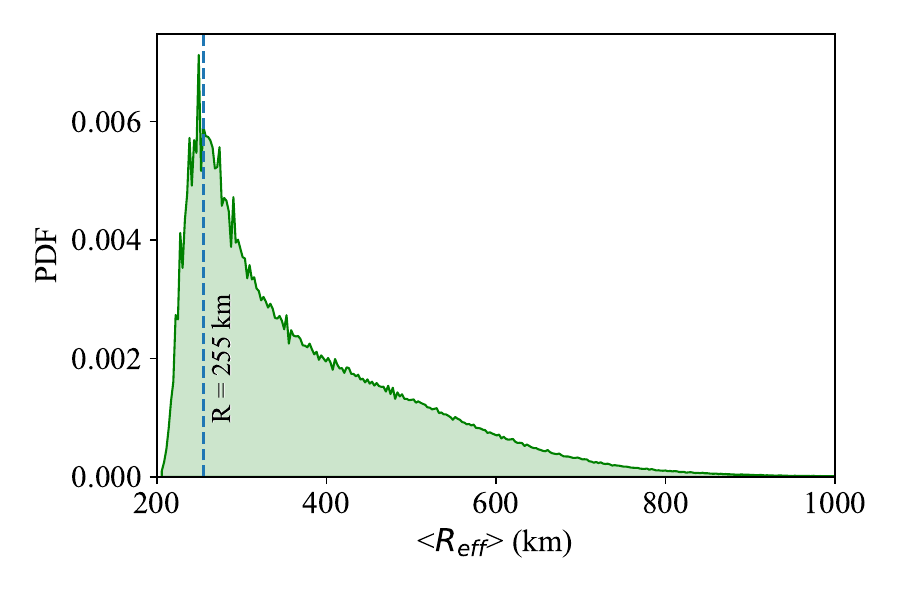}
         \end{subfigure}\\[3pt]
         \begin{subfigure}[b]{0.4\textwidth}
           \centering
           \includegraphics[width=\linewidth]{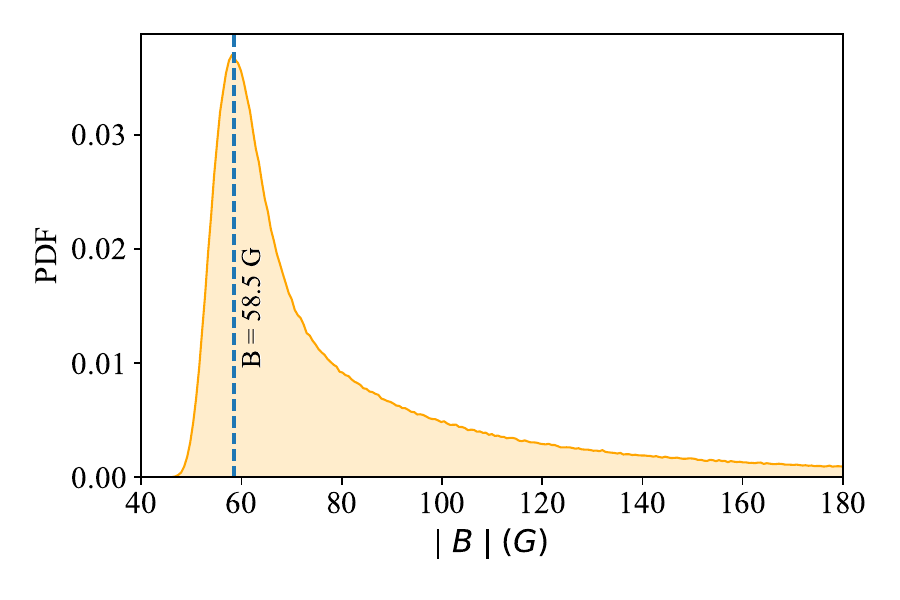}
         \end{subfigure}
         \caption{(Top) Probability density function of the effective radius of the magnetic elements investigated in this work. The vertical dashed line indicates the most frequent value, which is R=255 km. (Bottom) Probability density function of the mean magnetic field of the magnetic elements investigated in this work. The vertical line indicates the most frequent value of the magnetic field in the magnetic elements. Here the magnetic field values correspond to the mean values within the area of the elements.}
         \label{fig:stats}
     \end{figure}

\section{Methods and Results}
The tracking of photospheric magnetic elements in a small portion $400\times400$~arcsec$^2$ at disk center is done using the SWAMIS code \citep[][]{2007ApJ...666..576D}. The code searches for magnetic “blobs” above $B=40$~G, with an area of at least $4$~pixels and that are identified in at least $3$ consecutive images. These stringent thresholds taken together ensure the reliability of the tracking information of the magnetic elements in the photosphere and resulted in more than 31 million magnetic elements tracked in total. Each subset of $40$~min was coaligned to correct for the solar rotation. However, it is worth noting that the change in the degree of polarization of the horizontal perturbations seen in Fig.~\ref{fig:1} cannot be ascribed to an effect of the tracking, as in this case it would be independent of the solar cycle. The tracking allows the position of each magnetic feature in time to be determined and from that to estimate its horizontal displacement and oscillations. In this study, the amplitude of the horizontal (kink) oscillations was taken as the standard deviation of the horizontal velocity along the equatorial (East-West) and North-South axis ($\sigma_x$ and $\sigma_y$, respectively see Fig.~\ref{fig:1} panel c). In each temporal window of $40$ minutes, the average values of $\sigma_x$ and $\sigma_y$ of all the magnetic elements identified in the FoV (field of view) are considered.  This is done over the entire 11-year period considered, resulting in an estimate of the average amplitude of the perturbations every three days (Fig.~\ref{fig:1} panel c). The statistical set of magnetic elements tracked is constituted by magnetic concentrations with an effective radius peaking at about $255$~km. A more in-depth characterisation of the flux tubes tracked in this work and used in the analysis, as well as their statistical properties is shown in Fig. \ref{fig:stats}. The same data were also used in \cite{2024A&A...687L..21B}.\\
    It is found that the velocity amplitude of the horizontal perturbations of photospheric magnetic flux concentrations is not constant, but instead undergoes a long-term variation that follows the progression of the solar activity cycle (panel c of Fig. 1). It is seen that the horizontal velocity amplitudes resulting from the buffeting action of the photospheric plasma flows are the lowest at the point of solar minimum, yet increase as the solar cycle develops. This behaviour points to a change in the driver of these perturbations on timescales consistent with that of the activity cycle. Furthermore, a polarisation of the direction of their perturbations is also found, with the amplitude of the horizontal perturbations in the equatorial direction increasing faster than their counterparts aligned along the North-South (N-S) direction, implying a preferential direction of the underlying driver. This can be seen in Fig. 1 (panel d), where the differences of the velocity amplitudes in the two perpendicular directions are plotted.  Once again, we stress that both the amplitude and the polarization are reduced at the solar minimum,. This fact can be also seen in Fig. \ref{fig:2}, where we plot the probability density functions of the horizontal oscillations of the magnetic elements at both the solar maximum and minimum. Here we see that despite at the solar maximum the distribution of the amplitude of oscillations is broader in both x and y direction, in the x direction aligned to the equator the peak of the distribution is shifted toward the right, while in the y direction remains in the same position.\\
    It is worth underlining here that, what is investigated is not the bulk velocity of the magnetic elements, but the oscillation amplitude with respect to that. This implies that, although the data are co-aligned before the analysis, residual co-alignment errors resulting in a slow trend in the velocity time series associated to a magnetic element would not impact the results anyway. 

\section{Discussions}
    The results presented here show that the horizontal perturbations of the magnetic elements, with typical periodicities of the order of minutes, are not only driven by local plasma processes (e.g. buffeting of the flux tubes due to granular convection) but also by a long-term process with a characteristic timescale consistent with the solar cycle (i.e. several years). In other words, in addition to the locally acting driver given by the photospheric buffeting, there exists a larger scale driver acting on much longer timescales, typical of the solar dynamo and activity cycles.  It is worth noting here that both the amplitude variation and the polarization are highest at the peak of the solar cycle and decrease at the minima. This excludes the possibility of an effect of co-alignment of the data, which would also manifest at the minima.\\
    The observed modulation in the dynamics of these small-scale magnetic features over the solar cycle appears somewhat reminiscent of the (so called) torsional oscillations \citep[][]{1980ApJ...239L..33H, 2002Sci...296..101V, 2013ApJ...767L..20H, 2021ApJ...908L..50G}, which are cyclic perturbations in the solar differential rotation profile. 
    At low latitudes, the equatorward drift of the torsional oscillation pattern follows that of the sunspot cycle. Although the details of the driving mechanism are still not fully understood, the period of oscillation and the radial phase lag \citep[at least at low latitudes, see e.g.,][]{2002Sci...296..101V} suggests that the torsional oscillations are driven by the oscillatory large magnetic field within the solar interior \citep[e.g.][]{2004A&A...416..775C,2005AN....326..218B, 2016ApJ...828L...3G, 2019ApJ...887..215P}. Extending a similar analysis at higher latitudes could provide useful insights in this regard, also in preparation for the polar observations of Solar Orbiter in the coming years.
    The key result of the present paper is that there are indications that this dynamical influence is not confined to larger scales, but it also affects smaller ones explored by HMI near the surface. In fact, our observations suggest that the large-scale magnetic fields can also influence the dynamics of much smaller-scale magnetic features. However, the precise mechanism behind this possible coupling remains unclear. One possibility is that convective flows that are perturbing these magnetic elements are subject to some weak modulation by the large-scale solar magnetic field. Arguably a more likely explanation for these observations is that (at least some of) the magnetic features tracked might be relics of active regions decay \citep{2005ApJ...635..659H, 2021A&A...647A.146S}, with roots deep enough to couple with the global scale magnetic field.
    As the underlying large-scale magnetic field varies across the solar cycle, it is plausible that we would then observe some signatures of this variation in the dynamics of these magnetic elements. In this regard, higher resolution observations from e.g. DKIST \citep{2020SoPh..295..172R} could provide important information in this context, allowing the study of the photospheric magnetic fields dynamics down to very small scales (below $100$ km).
    We note in this regard that the elements tracked in this study are mostly represented by magnetic fields at the limit of HMI resolution and therefore they could be rather considered mostly as network fields or magnetic fields associated with the emergence of active regions. Whatever the underlying explanation, the solar-cycle variations in the horizontal perturbations of the magnetic elements in this study suggest that the large-scale solar magnetic field has a surprisingly important dynamical influence upon the short time-scale evolution of small-scale magnetic fields, not only in the subsurface layers but also at photospheric heights and beyond. 

\section{Conclusions}
Thanks to the availability of long-term continuous observations of the Sun's magnetic field with the SDO/HMI spacecraft, we studied the long-term modulation of the dynamical properties of small-scale magnetic structures in the photosphere. We found a clear correlation with the solar activity cycle, suggesting a possible connection between the global scales of the Sun's magnetic field cyclic regeneration, namely the solar dynamo, and the dynamics of small-scale structures in quiet-Sun regions; a perfect example for a complex system. Interestingly, a time-dependent polarization of the velocity perturbations is also observed with time scales compatible to that of the solar dynamo, further complicating our understanding of the driving mechanism. Concluding, future efforts will be directed toward the investigation of the non-trivial coupling of different temporal and spatial scales thanks to upcoming missions and further advancements in simulations.

\begin{acknowledgements}
We wish to acknowledge scientific discussions with the Waves in the Lower Solar Atmosphere (WaLSA; \href{https://WaLSA.team}{www.WaLSA.team}) team, which has been supported by the Research Council of Norway (project no. 262622), The Royal Society (award no. Hooke18b/SCTM; \citealt{2021RSPTA.37900169J}), and the International Space Science Institute (ISSI Team 502). MB acknowledges that this publication (communication/thesis/article, etc.) was produced while attending the PhD program in  PhD in Space Science and Technology at the University of Trento, Cycle XXXIX, with the support of a scholarship financed by the Ministerial Decree no. 118 of 2nd March 2023, based on the NRRP - funded by the European Union - NextGenerationEU - Mission 4 ``Education and Research'', Component 1 ``Enhancement of the offer of educational services: from nurseries to universities'' - Investment 4.1 ``Extension of the number of research doctorates and innovative doctorates for public administration and cultural heritage'' - CUP E66E23000110001. 
    SJ acknowledges support from the Rosseland Centre for Solar Physics (RoCS), University of Oslo, Norway.
    DBJ acknowledges support from the Leverhulme Trust via the Research Project Grant RPG-2019-371. DBJ and SJ wish to thank the UK Science and Technology Facilities Council (STFC) for the consolidated grants ST/T00021X/1 and ST/X000923/1. DBJ also acknowledges funding from the UK Space Agency via the National Space Technology Programme (grant SSc-009). VF and GV are grateful to the  Science and Technology Facilities Council (STFC) grants ST/V000977/1 and ST/Y001532/1. They also thank the Institute for Space-Earth Environmental Research (ISEE, International
Joint Research Program, Nagoya University, Japan), the Royal Society, International Exchanges Scheme, collaboration with Greece (IES/R1/221095), India (IES/R1/211123) and Australia (IES/R3/213012) for the support provided.
\end{acknowledgements}

\bibliographystyle{aa}

\end{document}